# Fermions and bosons in nonsymmorphic PdSb$_2$ with sixfold degeneracy

Ramakanta Chapai[1#], Yating Jia[2#], W. A. Shelton[3], Roshan Nepal[1], Mohammad Saghayezhian[1], J. F. DiTusa[1], E. W. Plummer[1], Changqing Jin[2*], and Rongying Jin[1*]

[1]*Department of Physics and Astronomy, Louisiana State University, Baton Rouge, LA 70803, USA*

[2]*Institute of Physics; School of Physics, University of Chinese Academy of Sciences, Chinese Academy of Science, Beijing, 100190, China.*

[3]*Cain Department of Chemical Engineering, Louisiana State University, Baton Rouge, LA 70803, USA*

PdSb$_2$ is a candidate for hosting 6-fold-degenerate exotic fermions (beyond Dirac and Weyl fermions). The nontrivial band crossing protected by the non-symmorphic symmetry plays a crucial role in physical properties. We have for the first time grown high-quality single crystals of PdSb$_2$ and characterized their physical properties under several stimuli (temperature, magnetic field, and pressure). While it is a diamagnetic Fermi-liquid metal under ambient pressure, PdSb$_2$ exhibits a large magnetoresistance with continuous increase up to 14 T, which follows the Kohler's scaling law at all temperatures. This implies one-band electrical transport, although multiple bands are predicted by first principles calculations. By applying magnetic field along the [111] direction, de Haas-van Alphen oscillations are observed with frequency of 102 T. The effective mass is nearly zero (0.045m$_0$) with the Berry phase close to π, confirming that the band close to the R point has a nontrivial character. Under quasi-hydrostatic pressure (p), evidence for superconductivity is observed in the resistivity below the critical temperature T$_c$. The dome-shaped T$_c$ versus p is obtained with maximum $T_c^{max}$ ~2.9 K. We argue that the formation of Cooper pairs (bosons) is the consequence of the redistribution of the 6-fold-degenerate fermions under pressure.

[#]*These authors contributed equally*

[*]*Corresponding authors: Jin@iphy.ac.cn; rjin@lsu.edu*

The discovery of topological properties in condensed matter started a new era of physics. Many fermionic particles and phenomena predicted in high energy physics are now experimentally observed in topological materials such as Dirac, Weyl, and Majorana particles [1-3]. Their nontrivial topology gives rise to exotic physical properties, opening the door to future electronics with low-power consumption. The nontrivial topology results from crossings of conduction and valence bands. Depending on crystal symmetry, such crossings can result in degeneracy (g) with g=2, 3, 4, 6, and 8. It is known that g=2 corresponds to Weyl fermions and g=4 corresponds to Dirac fermions. These fermions have been extensively studied both in condensed matter physics and high energy physics [4-9]. The cases of g=3, 6, and 8 are of particularly interesting as they can only be found in condensed matter systems, having no high energy analogues as constrained by Poincare symmetry [10].

Based on the density functional theory (DFT), $PdSb_2$ is predicted to have g=6, where the degeneracy is stabilized by its non-symmorphic symmetry: $Pa\bar{3}$ [10]. In this space group, there arises 3-fold degeneracy, which is doubled by the presence of time-reversal symmetry resulting in *6-fold degeneracy*. For $PdSb_2$, the 6-fold degeneracy occurs at the time-reversal invariant R point at the corner of the Brillouin Zone. Unlike Dirac and Weyl bands with linear dependence with energy, the 6-fold degenerate bands at the R point have quadratic energy dependence [10]. However, non-Abelian Berry curvature is expected under special holonomy [10]. Although $PdSb_2$ has been structurally characterized in early 1960s [11], its physical properties remain unknown [12]. We have grown $PdSb_2$ single crystals, confirming its non-symmorphic structure. For the first time, we report its electrical and magnetic properties under various stimuli (temperature (T), magnetic field (H), and quasi-hydrostatic pressure (p)). Several interesting properties are discovered: (1) $PdSb_2$ is a diamagnetic metal with Fermi-liquid ground state under ambient pressure, (2) there is large positive magnetoresistance (MR) without sign of saturation up to 14 T, which obeys Kohler's scaling law, (3) there are de Haas-van Alphen oscillations with frequency of 102 T when field is applied along the [111] direction, revealing nearly zero effective electron mass and a nontrivial Berry phase, (4) it becomes superconducting under the application of p with the maximum superconducting transition temperature; $T_c^{max}$~2.9 K, and

(5) first principles calculations indicate the change of band structure under pressure, leading to the charge redistribution.

PdSb$_2$ single crystal growth and phase determination are described in Ref. [13]. All X-ray diffraction (XRD) peaks can be indexed under a Pyrite-type cubic structure (space group 205 ($Pa\bar{3}$)) with the lattice parameter 6.464 Å (Fig. S1 (a) [13]); consistent with the previously reported value [14]. The magnetization measurements were performed in a MPMS (*Quantum Design*). At ambient pressure, the electrical properties were measured using the standard four-probe technique in a PPMS-14 T *(Quantum Design)*. The resistivity under quasi-hydrostatic pressure was measured via the four-probe method using a diamond anvil cell as described in Refs. [13, 15-17]. DFT calculations were performed using a plane-wave based approach that incorporates the projected-augmented wave method within the Vienna *ab-initio* simulation package (VASP) [13, 18-21].

Figure 1(a) shows the temperature dependence of the electrical resistivity of PdSb$_2$ along the *bc* plane ($\rho_{bc}$). Note $\rho_{bc}$ decreases with decreasing temperature between 2-400 K with $\rho_{bc}$(300K)~75.9 μΩ cm and $\rho_{bc}$(2K)~6.6 μΩ cm, indication of good metallicity. At low temperatures, the data can be fitted with $\rho_{bc}(T)=\rho_0+AT^2$ below 60 K with $\rho_0$=6.4 μΩ cm and $A$=2.8 nΩ cm K$^{-2}$ [see inset of Fig. 1(a)]. The T$^2$ dependence of $\rho_{bc}$ in this non-magnetic system indicates the Fermi-liquid ground state with dominant electron-electron scatting at low temperatures.

PdSb$_2$ is diamagnetic resulting from atom core contributions (Fig. S1(c) [13]). Applying H either parallel or perpendicular to the current I, we find that there is positive MR. Figure 1(b) displays the transverse (H⊥I) MR$_{bc}$ at indicated temperatures. The MR$_{bc}$ increases with decreasing temperature. At a fixed temperature, the MR increases with increasing field without any sign of saturation, reaching 174% at 2 K and 14 T. For H//I, the MR$_{bc}$ is about one order of magnitude smaller (Fig. S2 [13]).

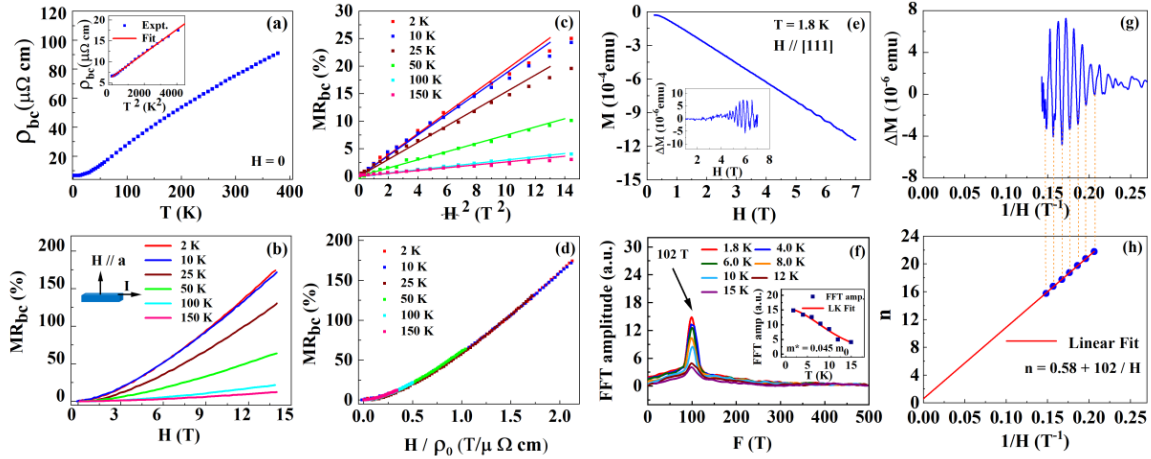

Fig.1. (a) Temperature dependence of the electrical resistivity ($\rho_{bc}$) of PdSb$_2$. Inset: $\rho_{bc}$ versus $T^2$ and fit with $\rho=\rho_0+AT^2$ below 60 K; (b) Transverse MR$_{bc}$ at indicated temperatures; (c) MR$_{bc}$ versus $H^2$ below 4 T; (d) Kohler plot at indicated temperatures; (e) Magnetization versus field at 1.8 K with H//[111]. Inset: $\Delta M$ after background subtraction; (f) FFT of oscillatory $\Delta M$ at indicated temperatures. Inset: temperature dependence of FFT amplitude and its fit to the Lifshitz-Kosevich formula (see text); (g) $\Delta M$ plotted as a function of $H^{-1}$; (h) Landau fan diagram constructed from $\Delta M$ at 1.8 K. The solid line is the fit of the data using the Onsager Equation.

Since the MR$_{bc}$ for H//I is much smaller than that for H⊥I, there is little contribution from spin scattering. The positive MR in a non-magnetic metallic system is attributed to the modification of electron trajectories due to the application of H. This effect is more significant at low temperatures in pure metallic systems where charge carriers effectively follow the cyclotron motion around the magnetic field [22-25]. By plotting MR$_{bc}$ versus $H^2$ in Figure 1(c), we find that the MR$_{bc}$ displays quadratic field dependence in the low-field regime (<3 T), as expected for conventional metals with a single band. At higher fields, MR$_{bc}$ gradually deviates from $H^2$ dependence, while continuously increasing. Such a non-saturating MR has been attributed to exotic mechanisms. For example, MR for the non-symmorphic Dirac semimetal Cd$_3$As$_2$ continuously increases up to 65 T, which is attributed to mobility fluctuations [26]. The MR of Cd$_3$As$_2$ also scales with H/$\rho_0$ [26], indicating an effective single-band transport. For PdSb$_2$, the MR$_{bc}$ data taken at different temperatures collapses into a single curve when plotting MR$_{bc}$ versus H/$\rho_0$ as shown in Figure 1(d). This implies that Kohler's law is valid for PdSb$_2$ as well, which is for a single-band transport scenario [25, 26]. Although calculations (Fig. 2 and [10]) indicate the multiband nature of PdSb$_2$, it appears that only one band plays a dominant role in electrical transport. In fact, the effective single-band transport in multiband systems is a common feature of topological materials [26].

While there is no sign for quantum oscillations when field is applied along the principle axes of crystals, we have observed such oscillations when the magnetization (M) is measured along the high-symmetry Γ-R direction. Fig. 1(e) shows H dependence of M along the [111] direction at 1.8 K; which exhibits oscillatory behavior with increasing H. The magnetization ΔM after subtracting a smooth background is shown in the inset of Fig. 1(e). The application of a fast Fourier transformation (FFT) to ΔM yields a single frequency F = 102 T (Fig. 1(f)) for all temperatures. The FFT amplitude decreases with increasing temperature. The temperature dependence of the FFT amplitude is used to determine the effective mass m* of electrons residing in this Fermi surface using the Lifshitz-Kosevich equation [27] $FFT(T) = \frac{A'\left(\frac{m^*}{m_0}\right)T}{Sinh(A'\left(\frac{m^*}{m_0}\right)T)}$ ($A' = \frac{2\pi^2 k_B m_0}{e\hbar \overline{H}}$, $m_0$ is the free electron mass, and $\overline{H}$ is the average magnetic field for our applied field range). The fit of this equation to our data yields $m^*=0.045m_0$. This value is close to what is obtained in $Cd_3As_2$ [26] and Weyl semimetal $BaMnSb_2$ [7].

The same frequency is obtained if ΔM is plotted as a function of 1/H as displayed in Fig. 1(g). Based on the observed oscillations, the Landau fan diagram can be constructed by assigning the minimum ΔM to n-1/4 [28], where n is the Landau level index. As shown in Fig. 1(h), n($H^{-1}$) can be described as n=0.58+102/H. This implies that $\Phi_{Berry}$=0.58*2π~1.16π [28], corresponding to a nontrivial Berry phase. The slightly larger $\Phi_{Berry}$ (compared to π) could be due to errors in determining the minimum ΔM, where there may be another oscillation with smaller amplitude and lower frequency. However, determining this low-frequency oscillation requires further measurement above 7 T. Additionally, from the Onsager relation F= ($\Phi_0/2\pi^2$)$S_F$ (($\Phi_0$ is the flux quanta), we can estimate the extreme Fermi surface area ($S_F$) normal to the [111] direction~0.98 $nm^{-2}$, corresponds to a small Fermi wave vector $k_F$~0.055 Å$^{-1}$.

To further understand the observed properties, the electronic structure is calculated using the DFT and shown in Figure 2(a). Consistent with previous results [10], there are 3 bands crossing the Fermi level ($E_F$) with contributions from both Pd and Sb. At the R point, there is a 6-fold degeneracy, marked by a circle in Figure 2(a). Three bands are congregating at the R point as the non-symmorphic symmetry protects this degeneracy. Since all of them are doubly degenerate due to the inversion symmetry, *6-fold degeneracy exists at the R point*. In addition, we find that Sb dominates the band structure around $E_F$ (Table S1 [13]).

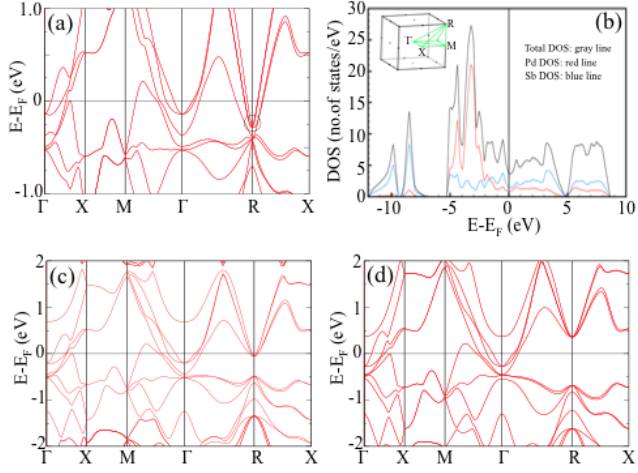

Fig.2. (a) Electronic structure near the Fermi level. (b) Density of states (DOS) from Pd (red), Sb (blue) and the sum. Inset: the first Brillouin zone of $PdSb_2$; (c-d) Band structure under p=17.66 GPa and 45.60 GPa, respectively.

The above analysis indicates that the topological states near $E_F$ at the $\Gamma$ and R points are primarily coming from Sb (between -0.37 eV and $E_F$). The DOS plots shown in Figure 2(b), show that there is a crossover at ~-1.5 eV where the site-projected Sb DOS overtakes the Pd site-projected DOS, indicating that on the more global scale that there is a larger contribution of Sb states near $E_F$ compared to the Pd states. Furthermore, the effective masses of the three bands at R point are much lighter than bands located at $\Gamma$ point. This indicates that the latter has higher DOS than that in the R point. Subsequently, the larger contribution of Sb than Pd to the DOS near $E_F$ suggests that the two lightest bands at R point are from Pd.

DFT calculations clearly show that there are both electron and hole bands across $E_F$. To figure out which band plays a dominant role in electrical transport, we measured the Hall resistivity ($\rho_H$) at various temperatures. As shown in Figure 3(a), $\rho_H$ exhibits linear field dependence with positive slope $R_H=\rho_H/H$ at all temperatures implying hole dominant charge carriers, and behave like single-band situation. The latter is in agreement with the Kohler's scaling behavior. The hole concentration estimated through $n = \frac{1}{eR_H}$, as shown in Figure 3(b), decreases with decreasing temperature, reaching $2.5 \times 10^{21}/cm^3$ at 2 K. The concentration of charge carriers is consistent with the good metallic behavior of $PdSb_2$ observed. According to $\mu_H = R_H/\rho$, the Hall mobility $\mu_H$ is calculated, which is shown in Figure 3(b). With decreasing temperature, $\mu_H$ increases, reaching 380 cm$^2$V$^{-1}$s$^{-1}$ at 2 K. This value of mobility is two orders of magnitude lower than that of Cd$_3$As$_2$ (8×10$^4$ cm$^2$V$^{-1}$s$^{-1}$) [8]. Since the resistivity of PdSb$_2$ is lower than that of Cd$_3$As$_2$, the lower $\mu_H$ is due to the higher n in PdSb$_2$.

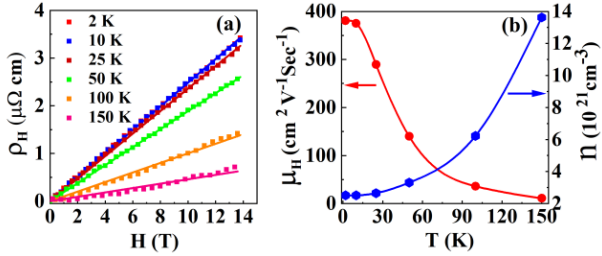

Fig.3. (a) Hall resistivity $\rho_H$ measured at different temperatures and corresponding linear fits (solid lines); (b) Temperature dependence of the Hall mobility $\mu_H$ (red) and carrier concentration n (blue).

Since the electrical transport properties of PdSb$_2$ are governed by carriers residing in one of the hole bands at the ambient pressure, would the electronic structure change under pressure? Figures 2(c-d) show the band structure under 17.66 GPa and 45.60 GPa respectively. While heavy bands ($\Gamma$ point) remain almost intact, the degenerate light bands (R point) are gradually pushed upward. Above~17.66 GPa, the light bands no longer cross $E_F$, as shown in Figure 2(d). This indicates that the DOS decreases with increasing pressure (Fig. S4 [13]). Moreover, a Bader charge analysis performed for p=0 and 45.60 GPa indicates that, under pressure, Pd gains charge while Sb loses charge, although the Bader volume is larger for Sb (Table S2 [13], [29-32]).

To probe the effect of the band shift with pressure, we performed the electrical resistivity measurements between 2-300 K under quasi-hydrostatic pressure up to 52 GPa. While the temperature dependence remains similar to that at ambient pressure, the magnitude of $\rho$ tends to decrease with increasing pressure. Upon increasing p, there is little change in $\rho$ in p$_1$ (=5 GPa) <p<p$_2$ (=35 GPa), as seen in Figures 4(a-b). Remarkably, above p$_2$, two dramatic features are observed: (1) the magnitude of $\rho$ increases with increasing p, and (2) there is sharp resistivity drop with p$\geq$41 GPa. Below, we quantitatively analyze these two features.

Under pressure, we find that $\rho$ continuously follows T$^2$ temperature dependence at low temperatures. We thus fit the low-temperature $\rho$(T) using the same formula as that for the ambient pressure case. Figure 4(c) displays both residual resistivity $\rho_0$ and coefficient $A$ under pressure. After initial drop, $\rho_0$ increases with increasing p when p≥2.6 GPa, with larger slope when p>$p_2$. The increase of $\rho_0$ with pressure may be because of the increased scattering with impurities in the squeezed environment. On the other hand, the coefficient $A$ initially decreases then increases after reaching the minimum at $p_2$. It further decreases after reaching the maximum at $p_3$~47 GPa. Within the Fermi-liquid theory, $A$ is proportional to DOS. The non-monotonic pressure dependence of $A$ implies the non-monotonic variation of DOS. This suggests that the shift of 6-fold-degenerate bands to higher energy by pressure (Figures 2(c-d)) involves charge transfer, thus changing the DOS. It should be pointed out that our XRD measurements under pressure show no evidence of any structural phase transition (Fig. S3 [13], [33]).

To confirm the change of DOS under pressure, we measured the pressure dependence of $\rho_H$ at 20 K. While the field dependence of $\rho_H$ is linear with positive slope as seen in Figure 3(a), the slope $R_H$ depends on p. Figure 4(d) shows the pressure dependence of n converted from $R_H$, which varies non-monotonically. There is a kink at $p_1$, a local maximum at $p_2$, and a minimum at $p_3$. The change of n reflects electron redistribution under pressure. Given the overall similar $\rho(T)$ profile under the wide range of pressure, the dominant hole carriers should be from the doubly degenerate band starting ~-0.6 eV at the X point that has hole pocket between the M and $\Gamma$ (Fig.2(a)).

We now turn to the sharp resistivity drop at low temperatures and high pressure which is shown in Figure 4(e). Note that there is obvious drop at 41 GPa. Although not reaching zero, the resistivity drop becomes sharper under higher pressure. This suggests that the system undergoes a superconducting transition. We determine the critical temperature $T_c$, at which $\rho$ reduced to 90% of its normal value. As plotted in Figure 4(f), $T_c$ initially increases then decreases after reaching maximum at $p_3$. Since $T_c$=1.2 K is expected at ambient pressure [12], there may be a superconducting dome as shown in Figure 4(f).

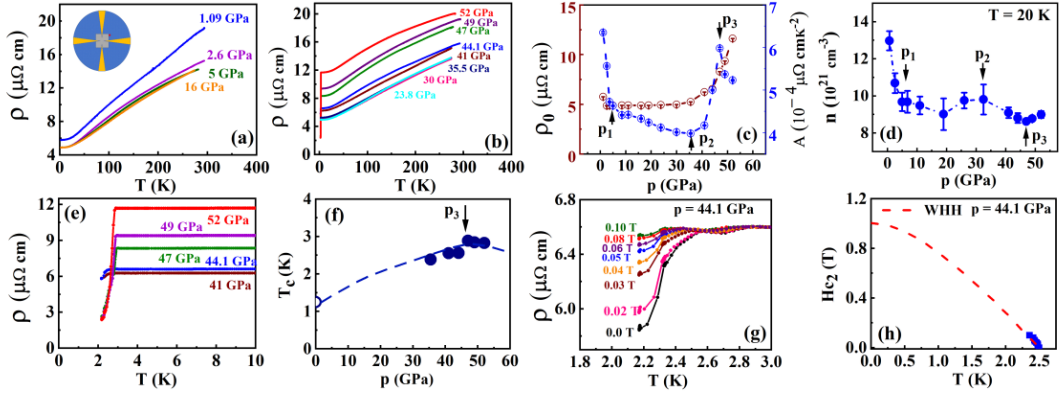

Fig.4. (a-b) Temperature dependence of ρ under pressure. Inset: schematic configuration of electrodes on the sample; (c) Pressure dependence of $\rho_0$ and $A$ (see text); (d) Pressure dependence of n at T=20 K; (e) Temperature dependence of ρ showing superconducting transition; (f) Pressure dependence of $T_c$, where the dashed line is to guide the eye based on information in Ref. [12]; (g) Temperature dependence of ρ under 44.1 GPa at indicated magnetic fields; (h) Temperature dependence of $H_{c2}$ at 44.1 GPa. The dashed line represents the WHH prediction based on the slope at $T_{c0}$~2.5 K.

To further check the origin of resistivity drop, we apply magnetic field at p=44.1 GPa. As shown in Figure 4(g), the resistivity drop is reduced upon the application of field, consistent with the scenario of a superconducting transition. The $T_c$ value at each field is plotted as the upper critical field $H_{c2}$ versus T [Fig. 4(h)]. Using the initial slope $dH_{c2}/dT|_{T = Tc0}$ ~ -0.58 T/K, we further estimate the $H_{c2}$ (T=0 K) ~1 Tesla using the Werthamer-Helfand-Hohenberg (WHH) approximation [34].

In view of pressure dependence of quantities shown in Figure 4, one may note that, at $p_3$, $T_c$ and $A$ reach the maximum while $n$ reaches minimum. This implies that strong electron-electron interactions (large $A$) in the environment of reduced carrier concentration favors superconductivity. Within the BCS theory, $T_c$ is determined by the Debye frequency, DOS, and electron-phonon coupling strength. Since $\rho(T)$ profile is unchanged over a wide temperature range, we may assume that there is no dramatic change in Debye frequency with pressure. This again points to the electronic origin for superconductivity.

In conclusion, we have successfully synthesized high-quality PdSb$_2$ single crystals with the non-symmorphic symmetry. Our first principles calculations confirm that PdSb$_2$ hosts 6-fold-degenerate fermions with a degeneracy stabilized by the non-symmorphic symmetry. By applying field along the high symmetry [111] direction, de Haas-van Alphen oscillations with F=102 T are observed, corresponding to a band residing electrons with nearly zero effective mass ($0.045m_0$) and a nontrivial Berry phase ($1.16\pi$). In addition to large and non-saturated magnetoresistance under magnetic field, the application of quasi-hydrostatic pressure modifies the electronic structure and changes superconducting transition temperature with $T_c^{max}$~2.9 K at $p_3$~47 GPa. Interestingly, the normal-state resistivity under pressure behaves similarly to that at ambient pressure, and follows Fermi-liquid behavior at low temperatures prior to the entrance of superconducting state. At $T_c^{max}$, the coefficient $A$ shows the maximum as well, while the carrier concentration reaches minimum. This suggests that the formation of Cooper pairs (bosons) is the consequence of strong electron-electron interaction. Our discovery of nearly massless electrons with nontrivial Berry phase and superconductivity in a 6-fold-degenerate material offers a unique system to study the behavior of both fermions and bosons.

**Acknowledgements**

This material is based upon work supported by the US Department of Energy under EPSCoR grant DE-SC0012432 with additional support from the Louisiana Board of Regents. The high-pressure experiment at Institute of Physics, Chinese Academy of Sciences was supported by China NSF and MOST research projects.

# Supplemental Materials

# Fermions and Bosons in Non-symmorphic PdSb$_2$ with Six-fold Degeneracy

Ramakanta Chapai, Yating Jia, W. A. Shelton, Roshan Nepal, Mohammad Saghayezhian,

J. F. DiTusa, E. W. Plummer, Changqing Jin, and Rongying Jin

**Sample preparation and characterization**

Single crystals of PdSb$_2$ were grown via the flux method using excess antimony (Sb) as the flux in a molar ratio of Pd : Sb = 1 : 2.2. The starting material, Pd powder (99.9%, Alfa Aesar) and Sb powder (99.5%, Alfa Aesar), were mixed together and placed into an alumina crucible, which was then sealed in a quartz tube under a vacuum of ~ 10 millitorr. The tube was heated to 980 ℃ at 120 ℃/h in a furnace, held at 980 ℃ for 48 h. Temperature was lowered down to 700 ℃ with a rate of 3.5℃/h. After staying at this temperature for 80 h, it was naturally cooled down to room temperature. Single crystals with typical size ~ 3 × 2 × 1.5 mm$^3$ were obtained. As-grown crystals were examined by performing powder X-ray diffraction (XRD) measurements (single crystals were crushed into fine powder) using a *PANalytical* Empyrean X-ray diffractometer (Cu κ$_α$ radiation; λ = 1.54187 Å). All peaks can be indexed under a Pyrite-type cubic structure (space group 205 ($Pa\overline{3}$)) with the lattice parameter 6.464 Å.

**Crystal Structure and magnetization**

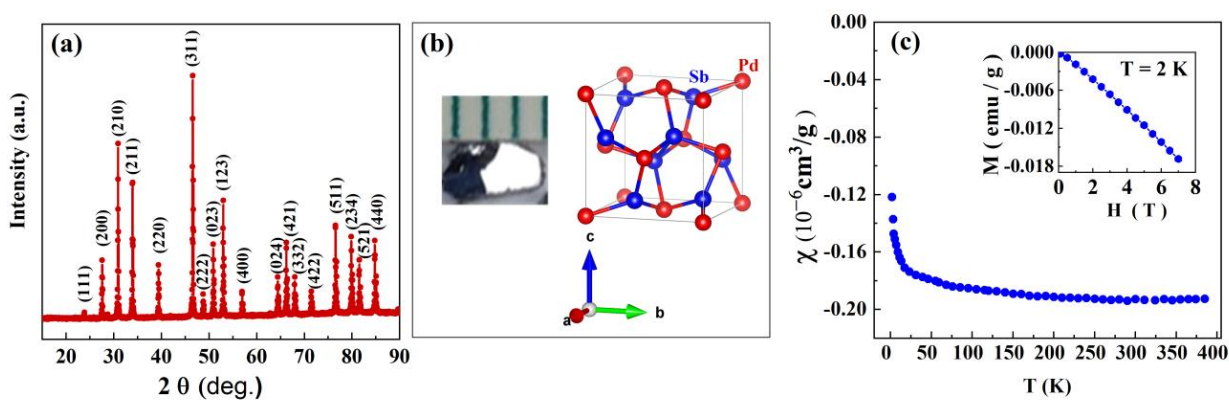

Fig.S1. (a) The room temperature X - ray diffraction pattern of PdSb$_2$ indexed in the $Pa\overline{3}$ structure. (b) A typical PdSb$_2$ single crystal (left) and the crystal structure (right). (c) Temperature dependence of the magnetic susceptibility measured by applying H = 1 kOe along the *a* axis. Inset: magnetization as a function of H at 2 K.

**Magnetoresistance of PdSb$_2$**

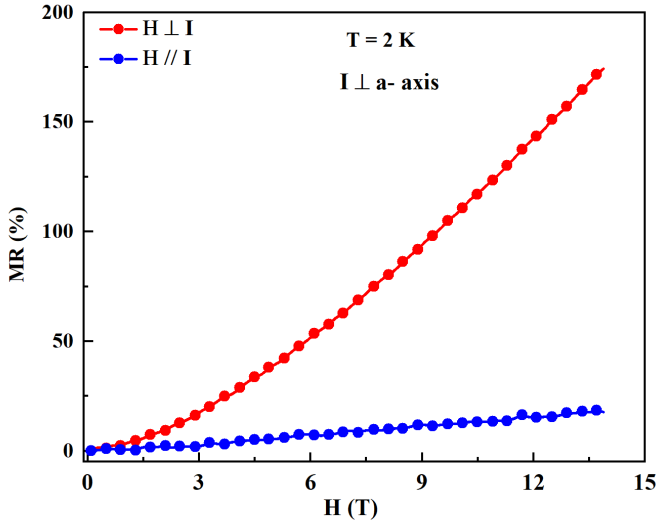

Fig. S2: Transverse magnetoresistance and longitudinal magnetoresistance of PdSb$_2$ at 2 K.

**Characterization under high pressure**

The electronic transport properties of PdSb$_2$ under high pressure were measured via four-probe method in a diamond anvil cell (DAC) made of CuBe alloy. Pressure was generated by a pair of diamonds with a 300 μm diameter culet. A gasket made of T301 stainless steel was pressed from a thickness of 250 μm to 20 μm, and drilled a center hole with a diameter of 200 μm. Fine cubic boron nitride (cBN) powder was used to cover the gasket to protect the electrode leads insulated from the metallic gasket. The electrodes are thin Au wires with a diameter of 18 μm. A 100 μm-diameter center hole in the insulating layer was used as the sample chamber. The dimension of the sample was about 70 μm × 70 μm × 5 μm, and NaCl powder was as the pressure transmitting medium. The pressure was measured via the ruby fluorescence method at room temperature before and after each cooling [1]. The diamond anvil cell was placed inside a Mag Lab system to perform the experiment. The temperature was automatically controlled by a program of the Mag Lab system. A thermometer was mounted near the diamond in the diamond anvil cell to monitor the exact sample temperature. Hall coefficients were measured via the Van der Pauw method.

For PdSb$_2$, in-situ high pressure angle-dispersive X-ray diffraction (ADXRD) experiments were performed using a symmetric Mao Bell DAC at Beijing Synchrotron Radiation Facility. The wavelength is 0.6199Å. The sample in DAC was fine powder ground from the single crystal, and a tiny ruby chip was used to measure the applied pressure. The two-dimensional image plate patterns obtained were converted to one-dimensional 2θ versus intensity data using the Fit2D software package [2]. Fig. S3 shows the ADXRD pattern for indicated pressure. As pressure gradually increases, the diffraction peaks of PdSb$_2$ shift steadily to higher angle, indicating the shrinking of lattices. Apart from the peak shift, the diffraction peaks become broader, but no new diffraction peaks emerges up to 44GPa.

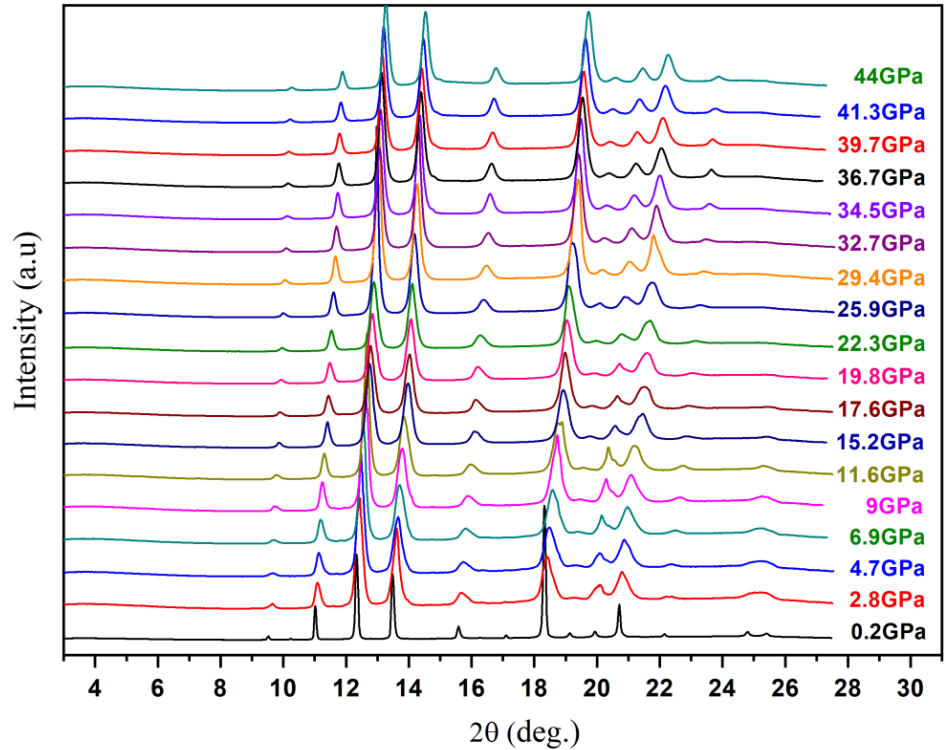

Fig. S3: Angle-dispersive X-ray diffraction of PdSb$_2$ under indicated pressure at room temperature.

**Band structure calculation**

The Density function theory (DFT) calculations were performed using a plane-wave based approach that incorporates the projected-augmented wave (PAW) method within the Vienna *ab-initio* simulation package (VASP). The generalized gradient approximation based Perdew-Burke-Ernzerhof (PBE) exchange-correlation functional was used [3]. The k-space integrations were carried out using a 16 × 16 × 16 Monkhorst-Pack special k-point mesh combined with the Methfessel-Paxton of order two integration methods with a Gaussian smearing factor of 0.2. We carefully studied the convergence of the simulations with respect to the energy cutoff and k-point where an energy cutoff of 328.9 eV and a 16 × 16 × 16 special k-point mesh were found to yield converged energies and forces.

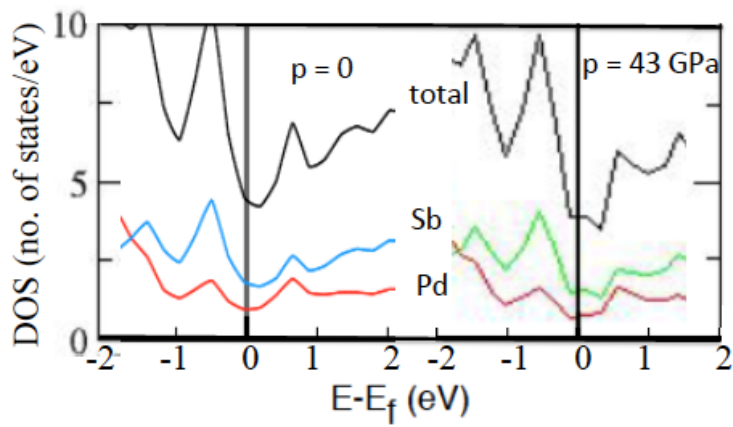

Fig. S4: Comparison of density of states from Pd (red) and Sb (green) and the sum for $PdSb_2$ between p = 0 and 43 GPa near $E_F$.

Table S1: Site contribution to the wave function at Γ and R points.

| Band | Γ | $E-E_f$ (eV) | Occupancy | R | $E-E_f$ (eV) | Occupancy |
|---|---|---|---|---|---|---|
| 183-Pd (4) | 0.548 | -0.48709064 | 1.02957679 | 0.539 | -0.41453974 | 1.05118226 |
| 183-Sb (8) | 0.248 | | | 0.316 | | |
| 184-Pd (4) | 0.548 | -0.48709064 | 1.02957681 | 0.538 | -0.41453973 | 1.05118226 |
| 184-Sb (8) | 0.248 | | | 0.234 | | |
| 185-Pd (4) | 0.068 | -0.36945667 | 1.02041925 | 0.560 | -0.38747627 | 1.03943165 |
| 185-Sb (8) | 0.472 | | | 0.304 | | |
| 186-Pd (4) | 0.068 | -0.36945667 | 1.02041924 | 0.560 | -0.38747627 | 1.03943164 |
| 186-Sb (8) | 0.472 | | | 0.304 | | |
| 187-Pd (4) | 0.063 | -0.13971847 | 0.09086159 | 0.071 | -0.28627859 | 0.78069372 |
| 187-Sb (8) | 0.463 | | | 0.434 | | |
| 188-Pd (4) | 0.063 | -0.13971847 | 0.09086159 | 0.072 | -0.28627859 | 0.78069372 |
| 188-Sb (8) | 0.463 | | | 0.434 | | |
| 189-Pd (4) | 0.061 | -0.13971847 | 0.09086082 | 0.072 | -0.28627825 | 0.78069220 |
| 189-Sb (8) | 0.465 | | | 0.434 | | |
| 190-Pd (4) | 0.061 | -0.13971847 | 0.09086079 | 0.072 | -0.28627825 | 0.78069220 |
| 190-Sb (8) | 0.465 | | | 0.434 | | |
| 191-Pd (4) | 0.004 | 1.03087423 | 0.00000000 | 0.072 | -0.28627808 | 0.78069147 |
| 191-Sb (8) | 0.432 | | | 0.435 | | |
| 192-Pd (4) | 0.004 | 1.03087423 | 0.00000000 | 0.073 | -0.28627808 | 0.78069146 |
| 192-Sb (8) | 0.432 | | | 0.434 | | |
| 193-Pd (4) | 0.280 | 2.57317772 | 0.00000000 | 0.284 | 2.03564082 | 0.00000000 |
| 193-Sb (8) | 0.374 | | | 0.360 | | |
| 194-Pd (4) | 0.280 | 2.57317772 | 0.00000000 | 0.285 | 2.03564091 | 0.00000000 |
| 194-Sb (8) | 0.374 | | | 0.359 | | |
| 195-Pd (4) | 0.279 | 2.57317772 | 0.00000000 | 0.284 | 2.03566612 | 0.00000000 |
| 195-Sb (8) | 0.375 | | | 0.360 | | |
| 196-Pd (4) | 0.279 | 2.57317775 | 0.00000000 | 0.285 | 2.03566632 | 0.00000000 |
| 196-Sb (8) | 0.375 | | | 0.359 | | |
| 197-Pd (4) | 0.296 | 2.61546236 | 0.00000000 | 0.285 | 2.03569197 | 0.00000000 |
| 197-Sb (8) | 0.379 | | | 0.360 | | |
| 198-Pd (4) | 0.296 | 2.61546238 | 0.00000000 | 0.284 | 2.03569216 | 0.00000000 |
| 198-Sb (8) | 0.379 | | | 0.361 | | |

Table S2: Bader charge analysis of PdSb$_2$ for p = 0, and 45.60 GPa [4-7].

| | p =0 | p = 45.83 GPa |
|---|---|---|
| Pd | 16.5821 | 16.8814 |
| Sb | 14.7089 | 14.5593 |
| Charge difference | 1.8732 | 2.3221 |